\documentclass{sf2a-conf2016}
\usepackage{graphicx}
\usepackage{hyperref}
\usepackage[]{natbib}  
\usepackage{epstopdf}
\usepackage{amssymb}
\usepackage{verbatim}

\def\BibTeX{{\rm B\kern-.05em{\sc i\kern-.025em b}\kern-.08em
    T\kern-.1667em\lower.7ex\hbox{E}\kern-.125emX}}
\bibpunct{(}{)}{;}{a}{}{,}  %%%%%%%%%%%%%  A&A bibliography style
\newcommand{\kms}{{\mathrm{km~s^{-1}}}}
\newcommand{\kpc}{{\mathrm{kpc}}}

\def\lya{\mbox{Ly$\alpha$}}
\def\lymana{\mbox Lyman-$\alpha$}
\begin{document}

\TitreGlobal{SF2A 2016}

%%-----------------------------------------------------------------
%%      the top matter
%%

\title{WEAVE-QSO: A Massive Intergalactic Medium Survey for the William Herschel Telescope}

\runningtitle{The WEAVE-QSO Survey}

\author{M. M. Pieri}\address{Aix Marseille Univ, CNRS, LAM, Laboratoire d'Astrophysique de Marseille, Marseille, France}\thanks{E-mail:matthew.pieri@lam.fr}

\author{S. Bonoli}

\author{J. 	Chaves-Montero}

\author{I. P{\^a}ris}

\author{M. Fumagalli}

\author{J. S. Bolton}

\author{M. Viel}

\author{P. Noterdaeme}

\author{J. Miralda-Escud{\'e}}

\author{N. G. Busca}

\author{H. Rahmani}

\author{C. Peroux}

\author{A. Font-Ribera}

\author{S. C. Trager}

\author{The WEAVE Collaboration}

%\author{J.-P. Author2}\address{Institute XYZ, 1299 City, OtherLand}

%% IF Author3 has the same affiliation than Author1:
%\author{C.\,E. Author3$^1$}

%% IF Author3 has its own affiliation:
%\author{C.\,E. Author3}\address{Dept. of Chess, University of Games, 35101 Las Vegas, Monaco} 

%% IF Author3 has two affiliations, the one of Author1 and a second one:
%\author{C.\,E. Author3$^{1,}$}\address{Dept. of Chess, University of Games, 35101 Las Vegas, Monaco} 

%% Keep this line, even if the page will be settled afterwards.
\setcounter{page}{237}

%%-----------------------------------------------------------------

\maketitle

%%-----------------------------------------------------------------
%%        The abstract
%% 
%%  Warning!  within the abstract:
%%  - do not use macros. 
%%  - do not use commands like: \cite, \citet, \citep ... etc.

\begin{abstract}
In these proceedings we describe the WEAVE-QSO survey, which will observe around 400,000 high redshift quasars starting in 2018. This survey is part of a broader WEAVE survey to be conducted at the  4.2m William Herschel Telescope. We will focus on chiefly on the science goals, but will also briefly summarise the target selection methods anticipated and the expected survey plan.

Understanding the apparent acceleration in the expansion of the Universe is one of the key scientific challenges of our time. Many experiments have been proposed to study this expansion, using a variety of techniques. Here we describe a survey that can measure this acceleration and therefore help elucidate the nature of dark energy: a survey of the \lymana\ forest (and quasar absorption in general) in spectra towards $z>2$ quasars (QSOs). Further constraints on neutrino masses and warm dark matter are also anticipated. The same data will also shed light on galaxy formation via study of the properties of inflowing/outflowing gas associated with nearby galaxies and in a cosmic web context. Gas properties are sensitive to density, temperature, UV radiation, metallicity and abundance pattern, and so constraint galaxy formation in a variety of ways. WEAVE-QSO will study absorbers with a dynamic range spanning more than 8 orders of magnitude in column density, their thermal broadening, and a host of elements and ionization species. A core principal of the WEAVE-QSO survey is the targeting of QSOs with near 100\% efficiency principally through use of the J-PAS ($r < 23.2$) and Gaia ($r \lesssim 20$) data.

\end{abstract}

%% Insert the keywords (to appear in the ADS indexing)
%% Keywords must be separated by a comma

%%-----------------------------------------------------------------

\section{Introduction}
%%---------------------
The  WEAVE  is a new multi-object survey spectrograph for the 4.2m William Herschel Telescope (WHT) with 1000 fibres over a 3.1 deg${^2}$ field of view. The WEAVE spectrograph offers two possible resolutions R=5000 and R=20000 \citep{2012SPIE.8446E..0PD,2014SPIE.9147E..0LD,doi:10.1117/12.2231078}.  WEAVE-QSO survey is designed to optimise quasar absorption science through the measurement of \lymana\ absorption and other intergalactic medium (IGM) absorbers. The science objectives form two pillars; probing cosmological parameters through measurements of baryon acoustic oscillations through quasar absorption, and a wider variety of IGM science and smaller-scale structure cosmology. The former is contingent on WEAVE-QSO's unrivalled number density of \lymana\ (\lya) forest quasars, the latter rests on unprecedented resolution and signal-to-noise massive spectroscopic survey.

Observations in the late 1990s unexpectedly showed that the expansion of the universe is accelerating at the present epoch. This is usually termed Ôdark energyÕ and one of the challenges of our time is determining the cause of this acceleration. These observations indicate that either our understanding of gravity is flawed on cosmological scales or that the majority of mass-energy in the Universe tends to push objects apart on these scales. One way to explore this phenomenon is to measure the expansion history of the Universe. There is a convenient standard ruler to achieve this, called Baryon Acoustic Oscillations (BAOs; e.g. Seo \& Eisenstein 2003). BAOs arise from acoustic waves in the early universe, and from these early perturbations large-scale structures were formed. Measuring the scale of BAOs in the distribution of large-scale structure at various epochs allows us to probe the expansion of the Universe.

The Baryon Oscillation Spectroscopic Survey (BOSS, \citealt{2013AJ....145...10D}) provides on average 17 QSOs deg$^{-2}$ (from 40 targets deg$^{-2}$) down to limiting apparent magnitude g=22 with single epoch data. This provides completeness of around 50\% in the critical redshift ranges $2.2<z<3.5$ \citep{2012ApJS..199....3R}. An average exposure of 45 min on a 2.5m telescope provides a median signal-to-noise in the \lya\ forest of around 2\AA. The extended version of BOSS (eBOSS, \citealt{2016AJ....151...44D}) is in the process of increasing the number density to 25 QSOs deg$^{-2}$ by adding 60,000 QSO at $z>2.1$, while it will also improve the S/N on a further 60,000 $z>2.1$ QSOs. In the process it is expected to obtain root-2 improvement on the BAO precision.

In a similar manner to SDSS and the survey for the Dark Energy Spectroscopic Instrument (DESI), the WEAVE-QSO survey will devote a small proportion of WEAVE fibres towards obtaining high redshift QSO spectra over a large footprint. However, we do not propose to duplicate the DESI \lya\ forest survey (which will occur on a similar timescale to WEAVE); rather we intend to use a complementary approach: WEAVE will narrow the redshift range and footprint providing a better BAO precision over this redshift range, and will open up the redshift and footprint for brighter QSOs for which WEAVE's higher spectral resolution will be most impactful. For both faint QSOs  ($r < 23.2$) and the bright QSOs ($r \lesssim 20$), we expect a near complete sample for their respective footprint and redshift cuts. This level of redshift and signal-to-noise selectivity is expected with near 100\% fibre efficiency for QSOs over from a single pass. A single pass is an efficient survey mode but it is also a necessity give a fibre reconfiguration time of approximately an hour for the WEAVE spectrograph. For comparison, DESI will take multi-pass approach to obtaining 90\% fibre efficiency on any $z>2.1$ QSOs. Excellent target selection is clearly a key element to the execution of the WEAVE-QSO survey. 

The steepness of the QSO luminosity function limits the viable number density of targets, but makes a survey of many thousands of square degrees desirable. Hence the WEAVE-QSO cannot dominate the fibre budget in any single field and must share fields with other programs. Key to this flexibility is the fact that a survey of structure in intervening absorbers such as this one is not sensitive to an uneven angular selection functions - i.e., we are free to target QSOs wherever they are available. High surface density of targets with sufficient signal to noise is desirable for cosmology though. Our proposed fields are predominately at high galactic latitudes in the NGC and as a result we expect to share fields with both the with the WEAVE Galactic Archaeology science  in both high- and low-resolution modes, and the WEAVE-LOFAR survey, both of which are also presented in these proceedings (Hill et al. {\it in prep} and \citealt{2016arXiv161102706S} respectively).

In the following we briefly set out a WEAVE-QSO survey science goals, and a brief summary of the current expected target selection and survey plan.

\section{Baryon acoustic oscillations in quasar absorption}

The measurement of clustering in diffuse intergalactic gas is an emerging method for measuring BAO and therefore dark energy. This measurement is performed using the \lya\ forest, a ÔforestÕ of absorption lines seen along the line-of-sight to distant QSOs caused by the intergalactic medium (IGM) on large scales. When the QSO has sufficiently high redshift ($z>2$), some of this forest of absorption falls in the optical window. Each QSO spectrum may, in principal, provide hundreds of megaparsecs of structure information along the line-of-sight in isolation, but moderate-resolution data, with low signal-to-noise, is limited by systematics on large-scales (related to uncertain continuum normalisation) and only provides structure information below $\le10h^{-1}$ Mpc (e.g. \citealt{2006ApJS..163...80M}). 

The BOSS survey overcame such limitations by building a sample of $\sim$150000 QSO \lya\ forest spectra over 5 years. BOSS provides, for the first time, a sufficiently high surface density of QSOs to characterise large-scale structures in the intergalactic gas between different lines of sight, thus reducing the impact of errors in any one line-of-sight \citep{2011JCAP...09..001S}. The \lya\ forest BAO feature has been detected  \citep{2013A&A...552A..96B,2013JCAP...04..026S} using a third of the expected BOSS QSO sample (Data Release 9: DR9). This provides the first measurement of the matter-dominated epoch where the expansion of the Universe was slowing. Recently, the autocorrelation measurement has been updated with twice the data \citep{2015A&A...574A..59D}, giving a 3\% precision measurement of BAO. The cross-correlation of quasars and the \lya\ forest has also been measured \citep{2014JCAP...05..027F}, generating an additional probe that is more sensitive to the angular diameter distance and with only limited correlation in errors with the forest autocorrelation. When combined, observations provide the most precise measurement to date of the Hubble parameter since the formation of the cosmic microwave background (CMB) and are in tension with the latest model based on CMB data from the Planck satellite at the 2.5$\sigma$ level. There are currently no physically motivated models \citep{2015PhRvD..92l3516A} to explain this tension and no known significant sources of systematic error. As a result we must obtain greater precision to either ease this tension or accept that we are driven to new physics.

\begin{figure}[ht!]
 \centering
 \includegraphics[width=0.50\textwidth,clip]{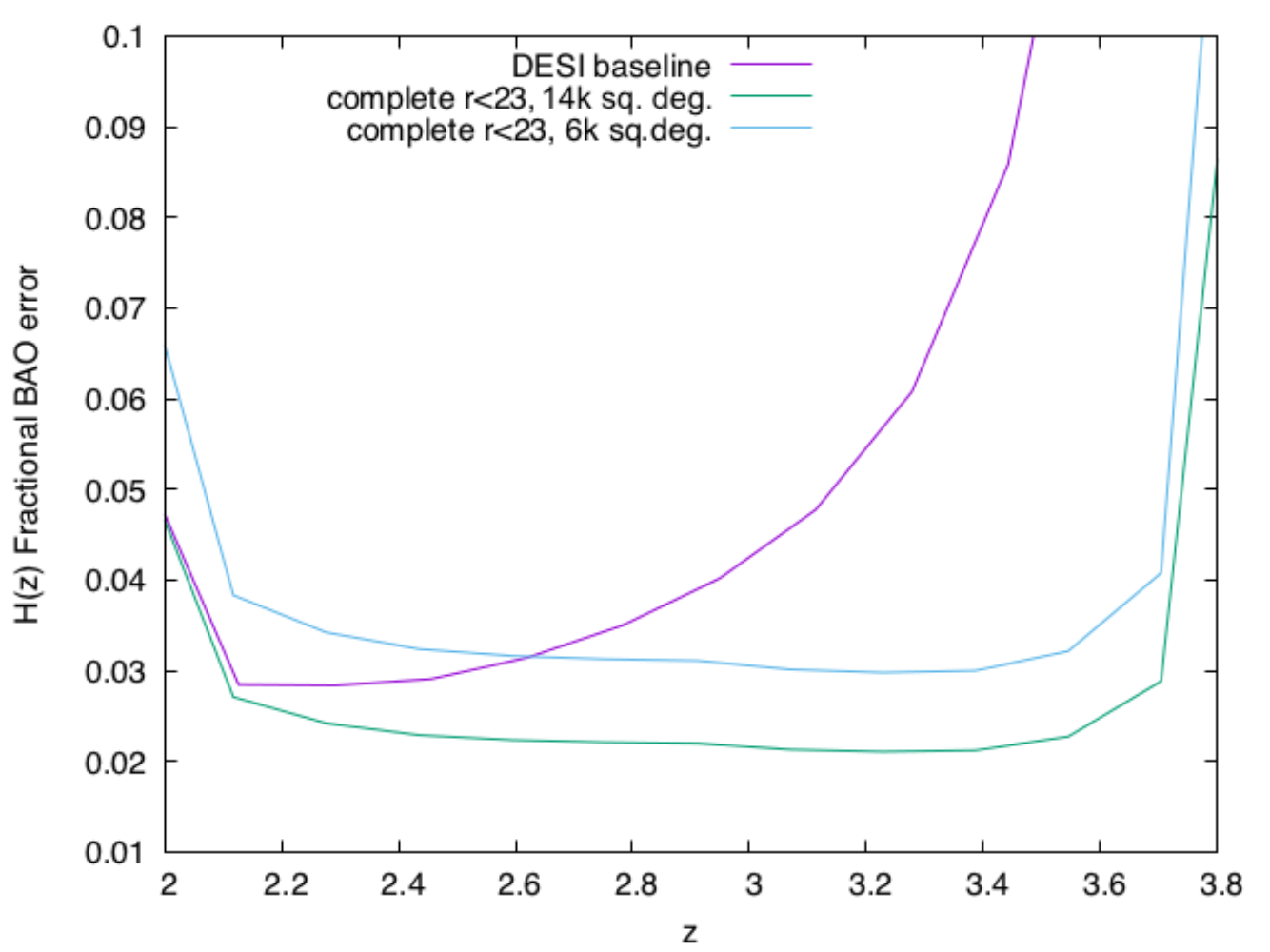}%      
 \includegraphics[width=0.50\textwidth,clip]{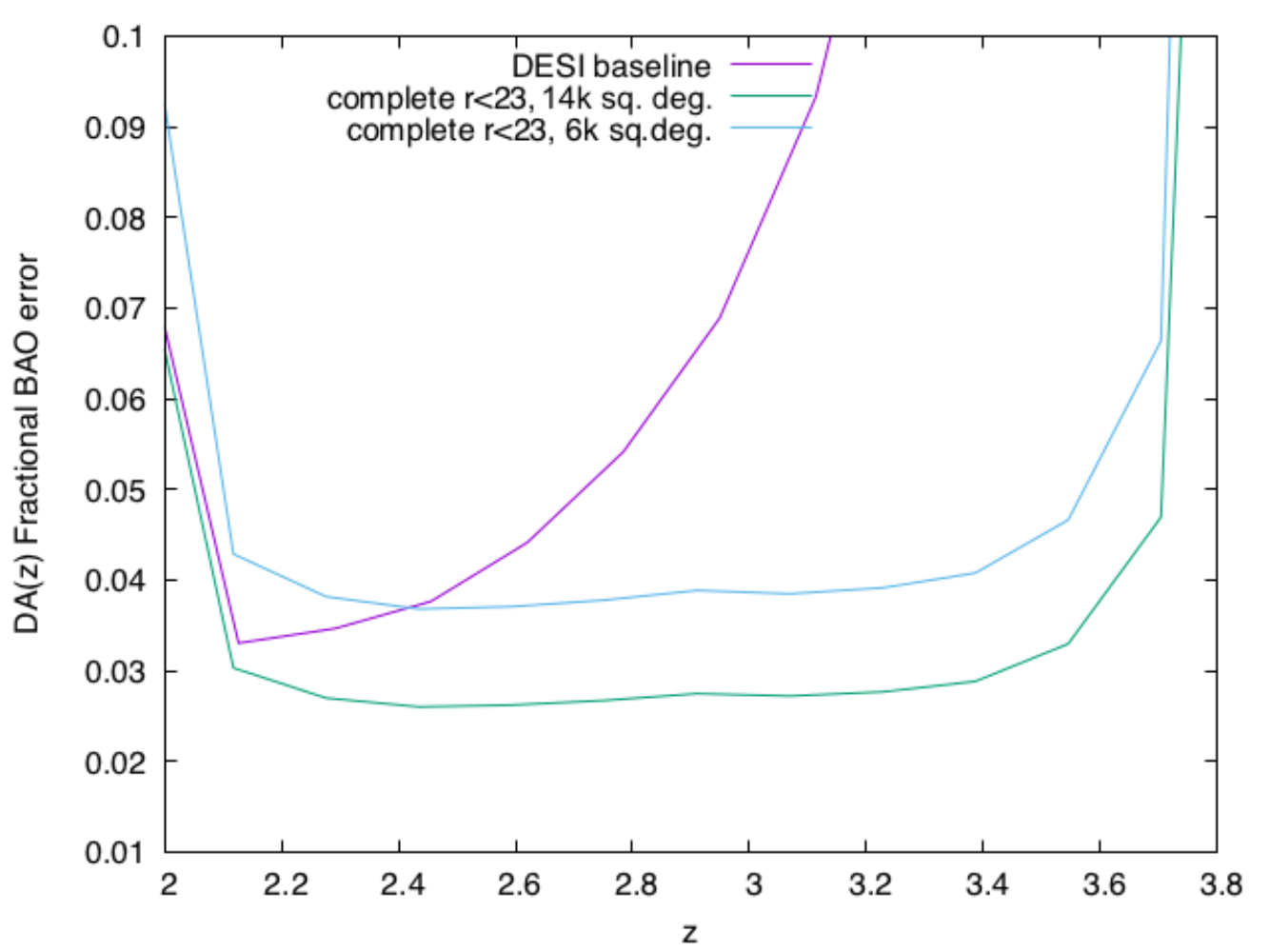}      
%% Note the ABSENCE of the extension .pdf  !
  \caption{Fisher forecasts of BAO precision expected for {\bf Left:} the Hubble parameter, and {\bf Right:} the angular diameter distance. This is shown for three survey scenarios (noting that DESI and WEAVE will produce spectra with similar signal-to-noise): the DESI baseline QSO absorption survey over 14,000 deg$^{2}$ to a depth of $r<23$, a complete QSO sample to  $r<23$ over $14,000 \rm{deg}^2$, and a complete QSO sample to  $r<23$ over 6000 deg$^{2}$. It is clear that for a survey complete to this depth, high BAO precision can be achieved for $z > 2.7$ which is also highly complementary to DESI. It should be noted that WEAVE-QSO also expected to survey $\Delta r \approx 0.2$ fainter. These forecasts were produced following the method of \citet{2014JCAP...05..023F}.}
  \label{BAO_proj}
\end{figure}

The WEAVE-QSO survey will measure BAO in the large-scale distribution of intergalactic gas  by taking advantage of the excellent completeness and purity of target selection provided by the Javalambre Physics of the Accelerating Universe Survey (J-PAS; \citealt{2014arXiv1403.5237B}). This targeting will be available for approximately $6000 \rm{deg}^{2}$ of the WEAVE-QSO survey, which can roughly be approximated by the overlap between J-PAS and SDSS survey boundaries shown in figure 3 of  \citet{2016arXiv161102706S}.
Figure~\ref{BAO_proj} shows that the greatest boost in BAO precision upon achieving a near complete sample is provided by a data with $z>2.7$. Despite the fact that the J-PAS targeting is only expected to cover $6000 \rm{deg}^{2}$ of the WEAVE-QSO survey, the survey is still expected to provide unrivalled intergalactic absorption BAO precision at $2.7<z< 3.5$ to approximately 0.5\% precision. Note also that the HETDEX survey (Hobby-Eberly Telescope Dark Energy Experiment,  \citealt{2011ApJS..192....5A}) is expect to achieve sub-percent precision on BAO with $2<z<4$ \lymana\ emitting galaxies.

\section{Other cosmology}
\label{othercosmo}

The \lya\ forest is currently providing the tightest limits on the total neutrino masses and on dark matter particle velocities (i.e. distinguishing between cold dark matter and warm dark matter models). A cosmological background of neutrinos is a prediction of the standard model and an important goal in modern cosmology is to detect it and constrain neutrino properties. The combination of cosmic microwave background data by Planck with the 1D \lymana\ flux power spectrum as measured from the SDSS-III survey \citep{2013A&A...559A..85P} has provided a tight upper limit of 0.15 eV at 2$\sigma$ confidence level on the total neutrino masses \citep{2015JCAP...11..011P}. The data consists of a set of about 15,000 quasar spectra from  BOSS DR9 among those that have high signal-to-noise ratios, which allowed to measure flux power in 12 redshift bins between z=2.2 and z=4.4 and at the scales between 0.002 $\kms$ and 0.02 $\kms$. These results show that we are on the verge of discriminating between the two possible neutrino hierarchies (inverted or normal). Furthermore, these findings have strong implications for particle physics experiments aiming at measuring neutrino masses or detecting the cosmic neutrino background.

WEAVE spectra in the low resolution mode have resolution nearly double that of SDSS (R=5000 as oppose to R=2000) and with this systematic effects will be better modelled and quantified. In particular, the metal contamination could be more quantitatively addressed; adding smaller scales to the \lymana\ flux power would allow also to probe further the neutrino induced suppression and its redshift and scale dependence (at the expenses of a more careful modelling from the theoretical side); a larger sample would result in similar statistical error bars on the flux power (though the statistical error contribution is still sub-dominant compared to the systematic ones); measuring the flux power to small scales $< 100\ \kms$) would allow to constrain the thermal state of the IGM more precisely; and this in turn will impact on the final constraints, since the thermal state history is usually marginalized over, and it is one of the main uncertainties in the modelling.

Regarding the coldness of dark matter, the \lya\ forest can also place tight constraints on the nature of dark matter by probing small-scales at relatively high redshift. Warm dark matter has been advocated in order to solve the small-scale crisis of the standard $\Lambda$CDM model presents, i.e., that there appear to be more low mass haloes, that are too dense and whose dynamical properties may be odds with observations. By possessing a non-negligible thermal velocity, warm dark matter significantly suppresses structure formation below a given scale. To constrain the particle mass it is mandatory to reach small scales (unlike neutrinos discussed above that have also an effect at larger scales) and also relatively high redshift, where non-linearities have had less time to erase primordial information in the power spectrum. The tightest constraints have been presented in \citet{2013PhRvD..88d3502V} who found a lower limit of $>3.3$ keV for a thermal warm dark matter relic, using a set of 25 $z>4$ Keck High Resolution Echelle Spectrometer and the Magellan Inamori Kyocera Echelle QSO spectra in combination with the SDSS 1D flux power at lower redshifts. This analysis shows that the flux power is consistent with very massive particles that are indistinguishable at this level from cold dark matter, while masses of 1-2 keV that are usually advocated to solve the small scale tensions present in the standard model are not supported by this data and analysis. WEAVE QSO spectra in the R$\sim$20000 are sufficient to measure the cut-off induced by the thermal state to a much more precise degree than the one which is usually measured by the few tens of high resolution spectra at similar redshifts that have been used so far. The possibility of probing other dark matter models such as wave (or fuzzy or quantum) dark matter (e.g. \citealt{2014NatPh..10..496S}; \citealt{2016arXiv161008297H}) present themselves.

\section{Cosmic web tomography and circumgalactic medium science}
%%-------------------------

A key facet of galaxy formation is its environmental context. In order to develop a thorough understanding of galaxy formation and intergalactic gas we must map structures in which they reside \citep{2001MNRAS.326..597P}. How do galaxy and gas properties differ in knots, filaments, sheets and voids and how do they evolve in these different environments? These are fundamental questions we are currently unable to address, although early attempts are being made in narrow, deep surveys \citep{2014ApJ...788...49L}. Instrumentation for next-generation ELT class telescopes is being developed with this specific science goal in mind.  WEAVE will pursue cosmic web mapping in three distinct programs; one wide and low resolution, another one  deeper with higher resolution, and a third making use of rare close groups of structure skewers.

Wide cosmic web mapping will make complete use of the whole deep-wide sample including faint QSOs. This sample allows IGM 3D mapping, where we expect to obtain $<15$ Mpc h$^{-1}$  resolution over this 6000 deg$^2$ footprint \citep{2016MNRAS.456.3610O}. This resolution will be sufficient to identify large-scale voids useful for void-counting and a measurement of the Alcock-Paczynski effect through void shape, both of which have demonstrated cosmological value in galaxy surveys (\citealt{2015PhRvD..92h3531P} and \citealt{2014MNRAS.443.2983S} respectively). While not sufficient to place the following studies of legacy astrophysics in a filamentary context, this will allow us to reconstruct of large-scale peaks and voids of structure and explore IGM properties and galaxy formation in the context of large-scale environment. J-PAS will provide a large number of \lymana\ emitting galaxies (around 300,000). In the best cases they will be detected with a redshift precision of $\Delta z=0.01$, this is also insufficient to resolve filamentary structure but may supplement low resolution tomography for a void/non-void separation. 

Higher resolutions can be obtained using the highly complementary galaxy called HETDEX. This survey provides galaxies identified in \lya\ emission (known as `\lymana\ emitters') at redshifts $z>2$, and so are highly complementary with this survey of \lya\ absorption. HETDEX begins this year (2016) and will cover 450 deg$^2$ within the proposed WEAVE survey footprint. This area will be filled with IFUs with a filling factor of 1/4.5 each with R=700 providing 1 Mpc precision on galaxy locations. Each IFU is 1' wide and is separated from others by $\sim$1'. Each IFU will be filled with \lymana\ emitters tracing structure and each can be thought of as providing a skewer of structure analogous to quasar absorption. 
WEAVE-QSO will reobserve $r<21$ QSOs in the HETDEX footprint, doubling the integration time and supplementing the S/N$>$5/\AA\ sample. \lya\ forest data will improve the mapping resolution by improving the filling factor of large-scale structure skewers, provide an alternative tracer for these structure to (aiding with the understanding of the use of LAEs as a tracer) and allow the possibility of studying the large-scale gaseous context of galaxy formation, feedback and the feeding of star formation. It is evident that mapping of the cosmic web with galaxies, while probing the \lya\ forest, will also generate a variety of legacy science of placing both galaxies and their circumgalactic media in a cosmic web context.

SDSS quality spectra have proven a rich source of information about metals associated with the optically thin gas of the \lya\ forest. This is despite the fact that the moderate resolution and S/N of such data render confident identifications of individual systems all but impossible. Progress was made by adapting the measurement of weak, distributed metal absorption through pixel-based techniques (e.g. \citealt{1995AJ....109.1522C}; \citealt{2003ApJ...596..768S}) to data of this quality. This has been demonstrated by directed measurements of a particular species \citep{2010ApJ...716.1084P} and blind searches that result from composite spectra produced by stacking forest lines \citep{2010ApJ...724L..69P, 2014MNRAS.441.1718P}. These composite spectra displayed unprecedented precision, providing the potential to measure gas column density, metallicity, elemental abundances pattern (resulting largely from stellar populations that give rise to them), and ionization fractions (arising due to UV background shape and intensity, density, temperature, and recombination time scale). \citet{2014MNRAS.441.1718P} showed that strong, blended forest lines on 138 $\kms$ scales are typically associated with circumgalactic regions of Lyman break galaxies such as those found in the KBSS survey \citep{2012ApJ...750...67R}. When such systems are stacked, the densities inferred force one to conclude that gas clumping on scales up to 30 pc and near-solar metallicity is seen. 

WEAVE data will enrich such \lya\ forest studies through larger samples with greater signal-to-noise and resolution. However, the combination of these measurements in combination with cosmic web mapping and \lymana\ emitter locations will be transformative for this science. The combination of  WEAVE and HETDEX surveys in particular will provide a wealth of information on the circumgalactic medium of \lymana\ emitting galaxies. These \lymana\ emitters are thought to reside in haloes of mass roughly an order of magnitude lower than the KBSS survey. Such haloes are more abundant and are predicted to dominate the volume fraction of the universe enriched by metals \citep{2007ApJ...662L...7P, 2012MNRAS.420.1053B}. The HETDEX IFUs will be sufficiently close to WEAVE quasars to probe CGM regions for approximately a third of WEAVE quasars over the 450 deg$^2$ footprint of HETDEX, this corresponds to approximately 4000 quasars with S/N/\AA\ $>5$ with full CGM information. This large statistical survey approach to studying the CGM of with \lymana\ emitters will complement ongoing focussed studies (e.g. \citealt{2016MNRAS.462.1978F, 2016A&A...587A..98W}) conducted with Multi-Unit Spectroscopic Explorer (MUSE).  This sample promises to be a rich source of gas inflow/outflow around galaxies at an epoch where quasar absorption spectra are a font of information (due primarily to a dense \lya\ forest observed in the optical window). This sample will allow the forest stacking techniques described above to be as discriminating as they are precise. Furthermore many Lyman limit systems (LLS) and damped \lymana\ (DLA) systems will be placed in a circumgalactic context boosting the science described below.

LLSs are optically-thick absorbers with column densities $>10^{17}$ cm$^{-2}$. Due to the characteristic absorption at the Lyman limit, they are easy to identify even in modest SN spectra, provided that the 912 \AA\ break in the system's rest-frame enters the spectral range of the survey ($z>3.01$ for WEAVE). LLSs are much more abundant than DLAs. At $z\sim3$, there are $\sim$5 times more LLSs than DLAs per unit redshift, and are impactful for both cosmological and galaxy evolution studies. By selecting quasars with narrow-band photometry, WEAVE will provide the first bias-free estimate of the number of LLSs and the mean free path of ionizing radiation in the $z\sim3$ universe, drastically reducing the uncertainties on current estimates the mean free path for ionizing photons. From the point of view of galaxy evolution, LLSs have recently been the subject of many studies, as they are believed to be associated to galaxy haloes. LLSs have in fact physical densities comparable to the density of halo gas ($<0.01\ {\rm cm}^{-3}$), and show a range of metal properties indicative of multiple gas phases in galaxy haloes (i.e. nearly pristine systems suggestive of gas accretion and highly enriched systems suggestive of outflows). The numbers of LLSs expected and the high resolution of the data allow the prospect of measurements of LLS bias and so their association with galaxies. Moreover, the resolution of WEAVE compared to sister surveys like DESI or BOSS allows the study of the metal enrichment of these absorbers, for the first time directly in a large survey. Furthermore WEAVE resolution offers the prospect of identifying more metal poor (or even metal free) DLAs, large samples of $z>3.5$ DLAs, and resolving metal velocity structure in both LLSs and DLAs.

\section{IGM thermal history }

The relationship between temperature and density in the intergalactic medium (IGM) is a fundamental quantity, which describes the physical state of the baryons in the early universe. Observations of the \lya\ forest in the spectra of high-redshift quasars are the primary approach used to obtain constraints on the IGM thermal history. Gas heating effects act as astrophysical Ônuisance parametersÕ \citep{2013PhRvD..88d3502V}. Improved measurements of the IGM thermal history will aid in refining these cosmological measurements (see Section~\ref{othercosmo}).

Photoionization heating impacts on the dynamical state of gas in the IGM by raising the gas pressure. This increases the scale over which small-scale structure is smoothed \citep{1998MNRAS.296...44G}. As there is a finite amount of time required for the gas to dynamically respond to a change in pressure, the precise amount of smoothing depends on when the gas was first heated and the hydrogen reionized. This ÔJeans smoothingÕ acts on larger scales (a few hundred proper kpc) relative to thermal (Doppler) broadening and smooths the gas in physical rather than velocity space \citep{2010MNRAS.404.1281P}. However, the pressure smoothing scale in the IGM at $2<z<3$ has not yet been measured precisely using line-of-sight \lya\ forest data. In principle, when combined with measurements of the instantaneous temperature, constraints on the IGM pressure smoothing scale from quasars may yield insights into the timing and duration of the reionization history.

The most widely used approach for studying the structure of the IGM in low to moderate resolution \lya\ forest spectra is the power spectrum of the transmitted flux. Combining the power spectrum with other statistics, such as the distribution of the transmitted flux, can further tighten constraints and help break parameter degeneracies \citep{2009MNRAS.399L..39V}. However, existing measurements are either based on data from the Sloan Digital Sky Survey, which consists of thousands of low to moderate resolution ($R\sim2000$), low signal-to-noise (S/N/\AA$\sim$5) spectra \citep{2006ApJS..163...80M}, or 10-m class telescope data consisting of several tens of high resolution (R$\sim$40000) high signal-to-noise (S/N/pixel$>$50) spectra (e.g. \citealt{2004MNRAS.347..355K}). The former data have achieved constraints on the \lya\ forest power spectrum that have small statistical error bars, while the latter, higher quality data are better suited to studying astrophysical effects on small scales. An independent sample of hundreds of moderate resolution (R$\sim$20000), moderate to high signal-to-noise (S/N/\AA$>$30) spectra at $2.4<z<3.4$ obtained with the WEAVE survey will probe an intermediate complementary regime. While resolving the thermal broadening requires higher resolution (R$\sim$40000) spectroscopy, R$\sim$20000 data will capture changes in the ionization state of the gas associated with evolution in the temperature-density relation. 

A precise measurement of the temperature-density relation will help to place further constraints on the tail end of He-II reionization at $z\sim3$ (e.g. \citealt{2009ApJ...694..842M}). He-II reionization is expected to result in large ($>30 {\rm Mpc} h^{-1}$) fluctuations in the ionization and thermal state of the intergalactic gas \citep{2009ApJ...694..842M}. These fluctuations are a potential systematic uncertainty in cosmological measurements. Ionization fluctuations will result in additional large-scale power in the three dimensional power spectrum of the \lya\ forest forest transmission \citep{2010ApJ...713..383W, 2014MNRAS.442..187G, 2014PhRvD..89h3010P}.
The WEAVE survey will provide a very large sample of low resolution ($R\sim5000$, cf. $R\sim2000$ for BOSS), moderate signal-to-noise spectra, with around 20 quasars per square degree to a limiting magnitude of $r=22.5$. 
\citet{2015MNRAS.447.2503G} estimate temperature fluctuations imprinted during He-II reionization at $z\sim3$ will impact at the 20-30\% level on the three-dimensional flux power spectrum at $k\sim0.02 {\rm Mpc}^{-1}$ for a BOSS-like survey with 15 quasars per deg$^{-2}$ and S/N/\AA$\sim$5. The WEAVE data will therefore be well-suited to this purpose, assuming that observational systematics (e.g. continuum fitting) and degeneracies with other astrophysical effects are well controlled and understood. This would provide a direct way to probe the expected patchy nature of He-II reionization, which is inaccessible with line-of-sight data alone. Similarly, exploring the impact of ionization fluctuations on the (even larger) scales associated with the mean free path for hydrogen ionizing photons, $k<0.01 \rm{Mpc}^{-1}$, will also provide independent constraints on the distribution of ionizing sources (i.e. quasars or star forming galaxies) at $2<z<3$ \citep{2014ApJ...792L..34P}.

\section{ Target Selection and Survey Size}
Our goal of near-100\% completeness and efficiency will be achieved by use of data from J-PAS. This is a narrow band imaging survey set to begin science verification in summer of 2016 and begin survey mode operations in mid-2017 and as such it leads the WEAVE survey approximately 1-2 year. These filters and provide a resolution of $R=50$. This survey will cover $>8500 {\rm deg}^2$ at high galactic latitudes, and will cover all trays over the course of 8 years. The boundary is dictated by a combination of observability and at Javalambre in Spain and limits on dust extinction. Given the location of the WHT, surveying the highest declination portion of this footprint is not possible. The observable portion of the J-PAS footprint can be approximated by its overlap with the SDSS DR8 imaging footprint, and amounts to approximately 6000 deg$^2$. This defines our deep-wide survey area. The study of QSO clustering and QSO science is already a core aspect of J-PAS science (e.g. \citealt{2012MNRAS.423.3251A}), and in coordination with the J-PAS team we have explored the identification of QSOs in J-PAS data.

\begin{figure}[ht!]
 \centering
 \includegraphics[width=0.45\textwidth,clip]{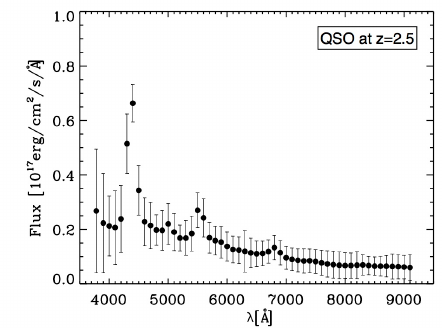}%      
 \includegraphics[width=0.45\textwidth,clip]{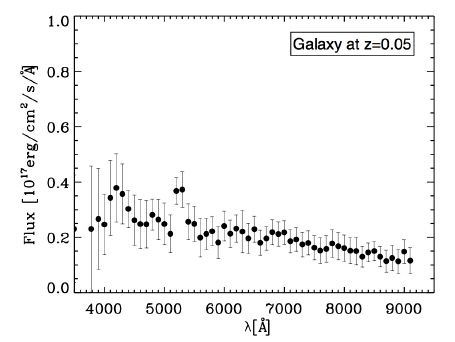}      
%% Note the ABSENCE of the extension .pdf  !
  \caption{An illustration of purity to galaxy interlopers. In both cases the points show perfect data and error bars indicating expected J-PAS uncertainty at a magnitude of r=23.2.  {\bf Left:} a $z=2.5$ QSO, exhibiting detectable \lymana, CIV and perhaps CIII in emission {\bf Right:} a mock z=0.05 galaxy spectrum that, in combination with a realisation of noise, is identified by a draft automated identification scheme as a z=2.5 QSO. The galaxy spectrum shows OIII emission that mimics CIV emission and a combination of nearby absorption lines which appears to mimic a weak \lymana\ emission line. However, the amplitude of these mimicking effects is both too weak with respect to continuum emission and one another to be mistaken for a z=2.5 QSO in a refined procedure. }
  \label{J-PASmocks}
\end{figure}

In order to test QSO identification in J-PAS we have built a library of 500 QSO templates to which we have added redshift dependent realisations of the \lya\ forest taken from BOSS cosmological mocks \citep{2015JCAP...05..060B}. These spectra then have the J-PAS filter response functions applied to them and a level as noise expected based on their assumed r-band magnitude. Recovery of high-z QSOs has then been tested through a visual inspection of QSOs and through automatic identification. In both cases the goal is to test for purity and completeness of QSOs with $2.2<z<2.9$ in J-PAS data, and to learn the faint limit of this identification.  This current test assumes that stars and galaxies have no impact on purity. We justify this due to the lack of strong emission lines that would pose a problem for identification in J-PAS data. As shown in Figure~\ref{J-PASmocks}, a galaxy and a z=2.5 QSO are easy to distinguish by eye despite our current automated methods identifying them as a z=2.5 QSO. Automated fitting algorithms are being developed, but visual inspection provides near perfect completeness and efficiency in current tests to $r<23$. If necessary visual inspection can be used to support automated approaches to achieve targeting of this quality.

\section{Summary}
%%--------------------

WEAVE-QSO is a massive spectroscopic survey of the intergalactic medium seen in absorption to background QSOs in absorption  set to begin in 2018. A deep-wide survey of \lya\ forest QSOs $r<23.2$ will be targetted using J-PAS data with 95\% completeness and purity within the 6000 deg$^2$ of overlapping footprint with the J-PAS survey. Throughout this footprint the WEAVE-QSO survey will take spectra of with $z>2.7$ QSOs down to this faint limit will constitute a signal-to-noise limit of S/N/\AA$>0.5$. This will be supplemented by QSOs with $2.1<z<2.7$ to a higher signal-to-noise limit for non-BAO science. In total this will provide 350,000 spectra of \lya\ forest QSOs.

This will be supplemented by survey of bright ($r<20$) QSOs with $z>2.1$  covering a further 4000 deg$^{-2}$ of Gaia targeted QSO giving S/N/\AA\ = 7. This provides 5 QSOs/deg$^{-2}$ or 16 per WEAVE field. By default this supplementary bright survey will be conducted in the low resolution mode, but a subset of this sample will be observed in the WEAVE high-resolution mode. This will be subject to fibre cross-talk constraints, sky brightness limits and redshift constraints based on science value and the limited wavelength coverage.

% Optional acknowledgements
% -------------------------
\begin{acknowledgements}
This work has been carried out thanks to the support of the A*MIDEX project (ANR- 11-IDEX-0001-02) funded by the ÒInvestissements dÕAvenirÓ French Government program, managed by the French National Research Agency (ANR).
\end{acknowledgements}

\bibliographystyle{aa}  % A&A bibliography style file (aa.bst)
%\bibliography{Pieri_WEAVE} % your references in file: Yourfile.bib

%
\end{document}